# High refractive index Fresnel lens on a fiber fabricated by nanoimprint lithography for immersion applications

Alexander Koshelev[†,1], Giuseppe Calafiore[†,1], Carlos Piña-Hernandez[1], Frances Allen[2,3], Scott Dhuey[2], Simone Sassolini[2], Edward Wong[2], Paul Lum[3], Keiko Munechika[1,2,*], Stefano Cabrini[2]

[1]aBeam Technologies Inc., 22290 Foothill Blvd. St. 2, Hayward, CA, 94541
[2]Molecular Foundry, Lawrence Berkeley National Laboratory, 1 Cyclotron Rd, Berkeley, CA 94720
[3]Biomolecular Nanotechnology Center, University of California, Berkeley, CA 94720
† G. Calafiore and A. Koshelev contributed equally to this work
*Corresponding author: km@abeamtech.com,

In this Letter we present a Fresnel lens fabricated on the end of an optical fiber. The lens is fabricated using nanoimprint lithography of a functional high refractive index material, which is suitable for mass production. The main advantage of the presented Fresnel lens compared to a conventional fiber lens is its high refractive index (n=1.69), which enables efficient light focusing even inside other media such as water or adhesive. Measurement of the lens performance in an immersion liquid (n=1.51) shows a near diffraction limited focal spot of 810 nm in diameter at the $1/e^2$ intensity level for a wavelength of 660 nm. Applications of such fiber lenses include integrated optics, optical trapping and fiber probes.

Lensed fibers have been used for a long time in integrated optics to improve coupling efficiency to optical chips [1]. They are also used as probes in various applications such as optical trapping [2], endoscopic imaging [3], optical coherence tomography [4] and glucose sensing [5]. The major limitation holding these applications back from further development is a poor focusing ability of the fiber lens upon immersion. The reason for this is the low refractive contrast between the lens and the surrounding medium. Conventionally, lensed fibers are made by laser ablation [6], polishing [7], wet etching [8] and focused ion beam micromachining [9]. In all these fabrication techniques, the lens structure is composed of the same material as the fiber, with a typical refractive index of 1.45. When such a lensed fiber probe is used inside a medium, for example water (n=1.33), the focusing ability is strongly reduced due to the small difference between the refractive indices.

In this Letter, we demonstrate a novel high refractive index Fresnel lens [10] fabricated directly on top of an optical fiber via Ultra Violet Nanoimprint Lithography (UV-NIL). NIL represents not only a low-cost, high-throughput nano-patterning technique with single-digit nanoscale resolution [11], but it also enables patterning of functional high refractive index materials, as previously demonstrated by our group and others [12-14]. Here, we demonstrate the fabrication of a Fresnel lens on a fiber by NIL of an imprint polymer [15] with a refractive index of 1.69 at 590 nm. We also demonstrate the focusing ability of the imprinted lensed fiber in immersion oil (n=1.51), which would be impossible using conventional fiber lenses. This demonstrates the significant advancement provided by such lensed fibers, particularly for integrated optics where an optical adhesive with matching index is commonly used to connect fibers to optical chips. For water-based immersion applications the refractive index contrast between the lens and the surrounding medium will increase by a factor of three (from Δn=0.12 to Δn=0.36). This leads to better light control in general, in particular in a tighter focal spot, which is advantageous for many applications. For example, in the case of optical trapping, a tighter focal spot allows trapping of smaller nanoparticles and increases the trapping force [16].

The imprinted lens presented here is fabricated following the process described in [17]. Focused Ion Beam (FIB) grayscale milling is adopted for the fabrication of the imprint mold. This provides a free-form, three-dimensional capability at the design stage. The choice of a Fresnel lens over the similarly performing aspherical lens was based on ease of fabrication. The Fresnel lens requires less material to be milled to create the mold. In addition, the Fresnel lens can have a larger working distance compared to the similar aspherical lens.

Lens design following an analytical approach does not account for waveguiding effects that occur inside the high aspect ratio structure. In order to correctly simulate such effects a three-dimensional FDTD (Finite-difference time-domain) code was used [18]. The design was iteratively optimized to minimize the focal spot diameter and side lobe intensity. The final lens design was targeted to operate inside a medium with a refractive index of 1.51 at a wavelength of 660 nm. Figure 1 shows the results of the simulation

The Fresnel lens is illuminated with a fiber mode field of 4 μm diameter. Even though most of the light is located within the central Fresnel ring, simulations show that the height and position of the other two rings have a significant effect on the focal spot size and side lobe intensity.

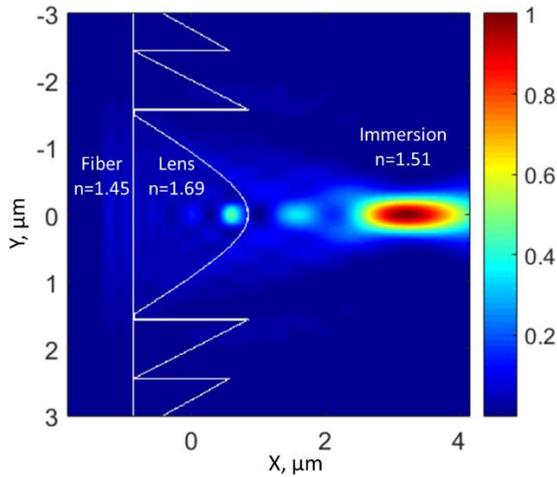

Figure 1. 3D FDTD simulation of the intensity distribution of 660 nm light propagating through the optimized Fresnel lens design. The focal spot diameter is 730 nm. A white line shows the lens contour.

The focal spot intensity profile obtained using the 3D FDTD simulation is shown in Figure 2. The simulated focal spot diameter is 730 nm at the $1/e^2$ intensity level. The intensity of the side lobes is small and does not exceed 1.3% of the maximum intensity.

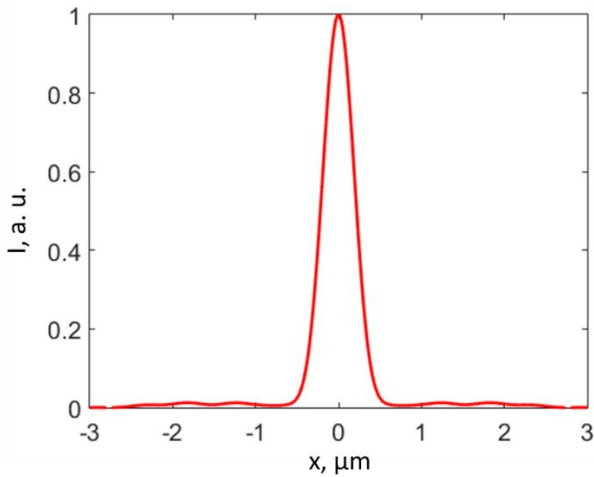

Figure 2. Simulated intensity profile of the Fresnel lens focal spot. The diameter of the focal spot is 730 nm at the $1/e^2$ intensity level.

The inherent disadvantage of the Fresnel lens is chromatic aberration. The simulated effect of chromatic aberrations is demonstrated in Figure 3. Wavelength shift results in an increase in the focal spot diameter. However, simulations show that the size of the focal spot is only within 5% of the optimal diameter over a spectral range of 85 nm around the design wavelength.

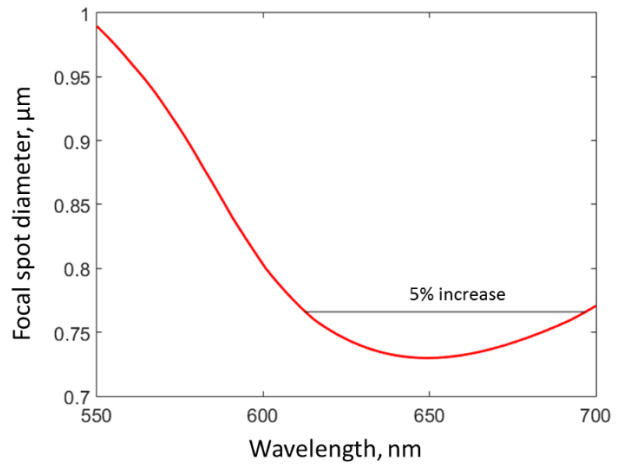

Figure 3. Simulation of the Fresnel lens focal spot diameter as a function of wavelength.

An imprint organic polymer with a refractive index of 1.69 at 590 nm, previously developed by aBeam Technologies, Inc. [15], was used for imprinting the lens. This functional polymer consists of low viscosity monomers and cross-linkers and it can be directly patterned by UV-NIL to create micro- and nanostructures with high transfer fidelity and accuracy. The material was specifically engineered to have a high optical transparency of more than 90% for a 5-μm thick film in the visible wavelength range.

The lens master mold was fabricated using a gallium FIB integrated into a Zeiss Orion NanoFab microscope. First and a second generation replicas of the master mold were created with a commercial UV-curable polymer (Ormocomp, Micro Resist Technology) [19]. The second replica, which has the same lithography tone as the master mold, is fabricated onto a transparent glass substrate to allow optical alignment between the fiber core and the Fresnel lens during imprinting. NIL was carried out using a custom built fiber imprinter, as described in Ref. [17]. A standard commercial single mode fiber (630-HP) was first immersed into the high refractive index material and then optically aligned to the lens mold. Sub-wavelength alignment is achieved, since the center of the diffraction spot can be determined with sub-diffraction accuracy. Once the fiber core is aligned to the imprint mold using the light through the fiber core as a guide, the fiber is brought into contact with the imprint mold and cured by irradiating with 405 nm light through the same fiber, after which it is released from the mold. Figure 4 shows a scanning electron microscopy (SEM) tilted-view image of the imprinted fiber. To the best of our knowledge, this is the purely organic polymer with the highest refractive index that has been imprinted and reported in literature.

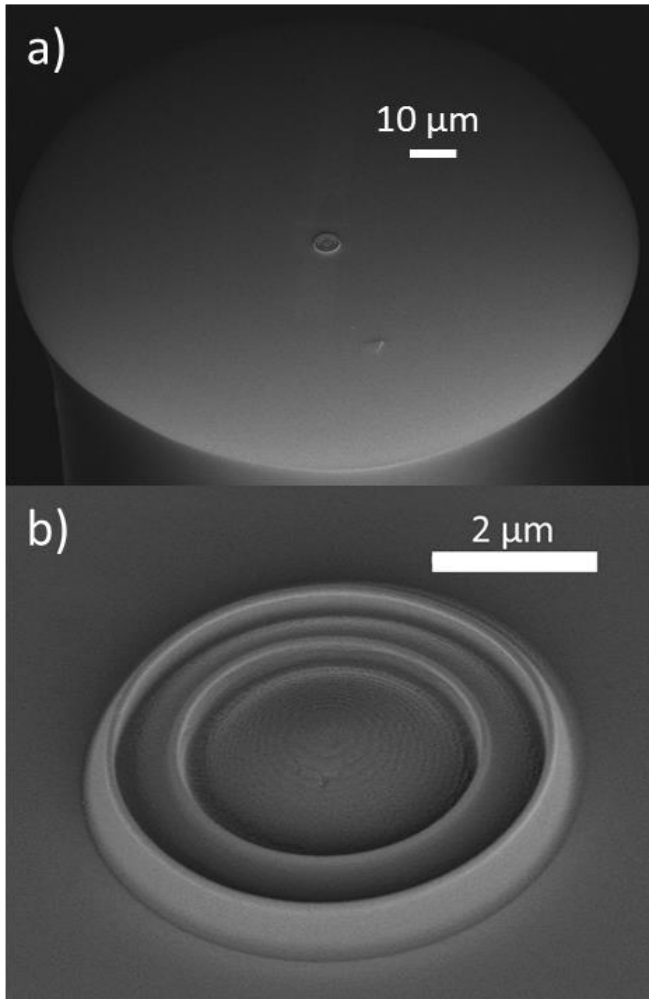

Figure 4. SEM image of the Fresnel lens imprinted onto an optical fiber. a) Low magnification view. b) high magnification view.

While fiber imprinting technique has been demonstrated previously for fabricating anti-reflective fiber coatings [20], and chemical sensors [21], our method presents several advantages including the use of a functional high refractive index imprint material, a 3D free-form imprint mold fabricated by FIB, and the capability to perform optical alignment of the fiber core to the mold.

To characterize the optical performance of the high refractive index Fresnel lens imprinted on a fiber, 660 nm laser light was coupled into the fiber and the focal spot was imaged in immersion oil (n = 1.51) using an oil immersion microscope objective (Nikon 100x, NA=1.3). For comparison, the same measurements were performed using a bare fiber. Figures 5a and b show camera images of the light intensity distribution of a bare and imprinted fiber, respectively. The intensities of the two images in Figure 5 were normalized so that the spatial distributions can be compared. The light spot diameter is much smaller in the case of the imprinted Fresnel lens, which confirms the accuracy of the lens design and functioning of our fabrication process. Figure 5c plots the cross-sectional intensity profiles obtained from figures 5a and 5b. The diameter of the focal spot measured for $1/e^2$ intensity was found to be 810 nm for the Fresnel lens. In comparison, a diameter equal to 4 µm was measured for the bare fiber, in agreement with the mode field diameter specified by the manufacturer of the fiber. The intensity of the side lobes is higher than expected from simulations, with the highest side lobe intensity being 9% as opposed to the simulated 1.3%. This might be explained by small fabrication inaccuracies introduced by the FIB and subsequent imprint. Nevertheless, the diameter of the focal spot is only 11 % larger than the simulated value. We note that using an objective for imaging may introduce additional aberrations. Thus, the actual fiber lens performance could be slightly better than the measured one.

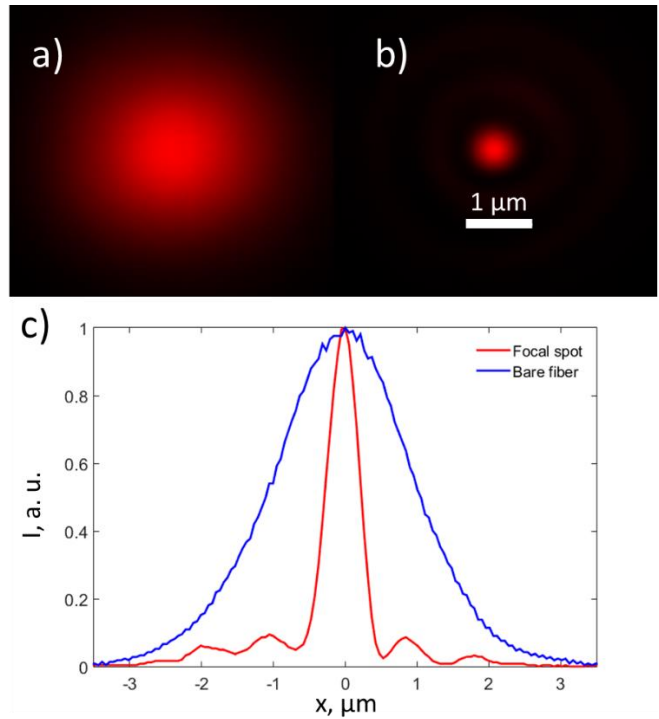

Figure 5. Light intensity distributions measured using an oil immersion microscope for a) A bare single mode fiber, and b) the focal spot of the imprinted Fresnel lens on the fiber. c) Corresponding intensity profiles. All intensities are normalized.

In conclusion, we designed, fabricated and tested a high refractive index Fresnel lens on a single mode fiber. The lens was fabricated using UV-NIL, which is a low cost and highly reproducible technique. Optical measurements confirm a focal spot with a diameter near that imposed by the diffraction limit achieved inside a liquid with a refractive index of 1.51. We believe that this NIL technique applied to a fiber can pave the way towards the inexpensive integration of other complex structures on fiber facets, such as vortex beam generators [22], near-field campanile probes [23], and beam shapers [24].

### Acknowledgement

This work is supported by the U.S. Department of Energy, Office of Science, Basic Energy Sciences, under Award Number DE-C0013109. Work at the Molecular Foundry was supported by the Office of Science, Office of Basic Energy Sciences, of the U.S. Department of Energy under contract no. DE-AC02- 05CH11231. The Zeiss ORION

NanoFab microscope is located at the Biomolecular Nanotechnology Center/QB3-Berkeley and was funded by an NSF grant from the Major Research Instrumentation Program (NSF Award DMR-1338139).

References
1. J. Cardenas, C. B. Poitras, K. Luke, L. W. Luo, P. A. Morton, and M. Lipson, IEEE Photonics Technology Letters **26**, 2380-2382 (2014).
2. H. Xin, Q. Liu, and B. Li, Scientific Reports **4**, 6576 (2014).
3. D. R. Rivera, C. M. Brown, D. G. Ouzounov, W. W. Webb, and C. Xu, Opt. Lett. **37**, 881-883 (2012).
4. B. H. Lee, E. J. Min, and Y. H. Kim, Optical Fiber Technology **19**, 729-740 (2013).
5. S. W. Harun, A. A. Jasim, H. A. Rahman, M. Z. Muhammad, and H. Ahmad, Sensors Journal, IEEE **13**, 348-350 (2013).
6. H.-K. Choi, D. Yoo, I.-B. Sohn, Y.-C. Noh, J.-H. Sung, S.-K. Lee, T.-M. Jeong, M. S. Ahsan, and J.-T. Kim, J. Opt. Soc. Korea **19**, 327-333 (2015).
7. T. Grosjean, S. S. Saleh, M. A. Suarez, I. A. Ibrahim, V. Piquerey, D. Charraut, and P. Sandoz, Applied Optics **46**, 8061-8067 (2007).
8. K. Taguchi, J. Okada, Y. Nomura, and K. Tamura, Journal of Physics: Conference Series **352**, 012039 (2012).
9. S. Cabrini, C. Liberale, D. Cojoc, A. Carpentiero, M. Prasciolu, S. Mora, V. Degiorgio, F. De Angelis, and E. Di Fabrizio, Microelectronic Engineering **83**, 804-807 (2006).
10. D. A. Buralli, G. M. Morris, and J. R. Rogers, Applied Optics **28**, 976-983 (1989).
11. C. Peroz, S. Dhuey, M. Cornet, M. Vogler, D. Olynick, and S. Cabrini, Nanotechnology **23**, 015305 (2012).
12. C. Pina-Hernandez, A. Koshelev, L. Digianantonio, S. Dhuey, A. Polyakov, G. Calafiore, A. Goltsov, V. Yankov, S. Babin, S. Cabrini, and C. Peroz, Nanotechnology **25**, 325302 (2014).
13. G. Calafiore, Q. Fillot, S. Dhuey, S. Sassolini, F. Salvadori, C. A. Mejia, K. Munechika, C. Peroz, S. Cabrini, and C. Piña-Hernandez, Nanotechnology **27**, 115303 (2016).
14. A. Pradana, C. Kluge, and M. Gerken, Opt. Mater. Express **4**, 329-337 (2014).
15. C. P. Hernandez, C. Peroz, and S. Cabrini, "Composition for resist patterning and method of manufacturing optical structures using imprint lithography," (Google Patents, 2016).
16. L. P. Ghislain, N. A. Switz, and W. W. Webb, Review of Scientific Instruments **65**, 2762-2768 (1994).
17. G. Calafiore, A. Koshelev, F. I. Allen, S. Dhuey, S. Sassolini, E. Wong, P. Lum, K. Munechika, and S. Cabrini, "Nanoimprint of a 3d structure on an optical fiber for light wavefront manipulation," in *ArXiv e-prints*(2016).
18. "Lumerical solutions inc. : A commercial-grade simulator based on the finite-difference time-domain method, vancoover, canada. Url https://www.Lumerical.Com/tcad-products/fdtd/.".
19. "Micro resist technology gmbh (www.Microresist.De)."
20. Y. Kanamori, M. Okochi, and K. Hane, Optics Express **21**, 322-328 (2013).
21. G. Kostovski, D. J. White, A. Mitchell, M. W. Austin, and P. R. Stoddart, "Nanoimprinting on optical fiber end faces for chemical sensing," (2008), pp. 70042H-70042H-70044.
22. P. Vayalamkuzhi, S. Bhattacharya, U. Eigenthaler, K. Keskinbora, C. T. Samlan, M. Hirscher, J. P. Spatz, and N. K. Viswanathan, Opt. Lett. **41**, 2133-2136 (2016).
23. W. Bao, M. Melli, N. Caselli, F. Riboli, D. S. Wiersma, M. Staffaroni, H. Choo, D. F. Ogletree, S. Aloni, J. Bokor, S. Cabrini, F. Intonti, M. B. Salmeron, E. Yablonovitch, P. J. Schuck, and A. Weber-Bargioni, Science **338**, 1317-1321 (2012).
24. T. Gissibl, M. Schmid, and H. Giessen, Optica **3**, 448-451 (2016).